\documentclass[preprint]{aastex}

\newcommand\las{\mathrel{\hbox{\rlap{\hbox{\lower3pt\hbox{$\sim$}}}\hbox{\raise2pt\hbox{$<$}}}}}
\newcommand\gas{\mathrel{\hbox{\rlap{\hbox{\lower3pt\hbox{$\sim$}}}\hbox{\raise2pt\hbox{$>$}}}}}
\newcommand{\degree}{^{\circ}}
\newcommand{\eg}{{\it e.g.\ }}
\newcommand{\ie}{{\it i.e.\ }}

\begin{document}

\title{The return of the Andromedids meteor shower}

\bigskip\bigskip

\author{Paul A. Wiegert, Peter G. Brown, Robert J. Weryk and Daniel K. Wong}
\affil{Dept. of Physics and Astronomy, The University of Western Ontario, London Canada N6A3K7}
\notetoeditor{\\CORRESPONDING AUTHOR:\\ Paul Wiegert\\Dept. of
Physics and Astronomy\\The University of Western Ontario\\London, Ontario\\Canada N6A3K7\\
pwiegert@uwo.ca\\Phone/fax: 1-519-661-2111x81327}

\slugcomment{\it Submitted to the Astronomical Journal Sep 22 2012}


\shorttitle{The Andromedids}

\begin{abstract}

The Andromedid meteor shower underwent spectacular outbursts in 1872
and 1885, producing thousands of visual meteors per hour and described as
`stars fell like rain' in Chinese records of the time \cite[]{kro88,
  nog95}. The shower originates from comet 3D/Biela whose
disintegration in the mid-1800's is linked to the outbursts, but the
shower has been weak or absent since the late 19th Century.

This shower returned in December 2011 with a zenithal hourly rate of
approximately 50, the strongest return in over a hundred years.  Some
122 probable Andromedid orbits were detected by the Canadian Meteor
Orbit Radar while one possible brighter Andromedid member was
detected by the Southern Ontario Meteor Network and several
single station possible Andromedids by the Canadian Automated Meteor
Observatory.

The shower outburst occurred during 2011 Dec 3-5. The radiant at RA
+$18\degree$ and Dec +$56\degree$ is typical of the `classical'
Andromedids of the early 1800's, whose radiant was actually in
Cassiopeia. The orbital elements indicate that the material involved
was released before 3D/Biela's breakup prior to 1846. The observed
shower in 2011 had a slow geocentric speed (16 km~s$^{-1}$) and was comprised
of small particles: the mean measured mass from the radar is $\sim5
\times 10^{-7}$~kg corresponding to radii of 0.5~mm at a bulk density
of $1000$~kg/m$^3$.

Numerical simulations of the parent comet indicate that the meteoroids
of the 2011 return of the Andromedids shower were primarily ejected
during 3D/Biela's 1649 perihelion passage. The orbital
characteristics, radiant, timing as well as the absence of large
particles in the streamlet are all consistent with
simulations. Predictions are made regarding other appearances of the
shower in the years 2000-2047 based on our numerical model.  We note
that the details of the 2011 return can, in principle, be used to
better constrain the orbit of 3D/Biela prior to the comets first
recorded return in 1772.

\end{abstract}


\section{Introduction} \label{section1}
\begin{quote}
``We can only hope that future perturbations will again switch the
  group [Andromedids] across our path, so that more can be learned of
  the processes at work related to the evolution and disintegration of
  the comet and how far they have progressed.'' \cite{oli25}
\end{quote}

In 1845/46, comet 3D/Biela was observed to be in the process of
fragmenting, a process which continued until the comet disappeared
entirely following its 1852 return. The break-up was followed by a
strong shower (the Andromedids), a shower first reported in 1798 and
displaying spectacular outbursts in 1872 and 1885 during which
thousands of meteors per hour were reported \cite[]{kro88,nog95}. The
shower has decreased in intensity since that time and since the early
20th century rates are less than a few per hour \cite[]{hawsouste59}.

Comet 3D/Biela was a Jupiter-family comet first discovered in March
1772 by J. L. Montaigne \cite[]{kro99}. This comet is significant for
being the first comet to be `lost' and acquire the `D' designation
instead of the usual `P' designation of periodic comets. It is also
among the first comets to be linked to meteor activity, as Weiss,
d'Arrest and Galle all independently reported the link between the
Andromedid shower and Biela's orbit \cite[]{kro88}.

According to \cite{jen06} and \cite{hawsouste59}, probable returns of the Andromedids date
back to roughly the mid-18th Century. Observations of the stream from
the mid-18th to mid-19th century report the time of maximum to be in
the first week in December, while the major storms of 1872 and 1885
occurred on Nov 27. Moreover, the few reports of strong activity from
the shower in the late-18th Century have times of maximum
progressively earlier, including Nov 24, 1892 and 1899; Nov 22, 1904
and finally Nov 15, 1940 when the last activity at a level of a few
tens of meteors per hour was noted. \cite{hawsouste59} describes
annual activity from the stream visible in the mid-20th century among
Super-Schmidt camera meteor data, but of weak intensity.  The
decreasing times of maximum reflect the change in the nodal longitude
of the 3D/Biela and the increasingly young trails encountered
in later years. From Jennisken's 2006 compilation (see his Table 6a),
the apparent radiant of the shower has moved in concert with the
changing date of maximum. The earliest showers, having maxima in
December/very late November had radiants near RA=$+20\degree$ and
Dec=$+50\degree$ to $+60\degree$ while the later showers (1850
onwards) and the storms of 1872/1885 had radiants of RA=$+27\degree$,
Dec=$+44\degree$. It is this latter era of the shower activity,
punctuated by the storms of 1872/1885 with the radiant in Andromeda
which gave the shower its modern name. In fact, the first measurements
of the shower showed the radiant to be in Cassiopeia, a feature of the
shower long recognized (e.g. \cite{oli25}). The change in the radiant
in concert with the date reflects the very different epochs of
ejection from 3D/Biela and its changing orbit through
different eras.

The link between the Andromedids and Biela was examined most recently
by \cite{jenvau07}. They studied the outbursts of the 19th century in
order to determine whether they were primarily the result of material
released during the splitting event or by the usual process of water
vapour sublimation, and conclude that ongoing fragmentation
particularly during the 1846 passage after the splitting event is most
likely responsible for the 1872 and 1885 outbursts.  Confining their
ejection epochs to perihelion passages of the comet after 1703, they
were able to model quite a number of appearances of the Andromedids
through to 1940. 

However, a few occurrences of observed showers did not appear
in their simulations: these may have been produced by perihelion
passages of the comet prior to 1703, the first perihelion they
consider. They did not examine possible appearances of the shower in
the current era.

Here we describe the first modern, strong return of the Andromedid
shower occurring during 2011 Dec 3-5.  Section~\ref{observations}
describes the observations of the meteor shower,
Section~\ref{simulations} outlines the simulations and
Section~\ref{results} presents a comparison of the two. Conclusions are drawn in
section~\ref{conclusions}.

\section{Observations} \label{observations}

The first indication of unusual activity associated with the
Andromedids in 2011 was made during post-processing of radar
measurements conducted by the Canadian Meteor Orbit Radar (CMOR) as part of a
program aimed at detecting brief shower outbursts. The Canadian Meteor
Orbit Radar is a multi-station, backscatter radar system operating at
29.85 MHz which is able to measure trajectories and speeds for
individual meteors. Details of the basic system can be found in
\cite{webbrojon04,jonbroell05,browerwon08} and \cite{browonwer10}. The original
CMOR system began orbital measurements in 2001 using a three station
setup. In mid-2009, CMOR was upgraded to higher transmit power (12 kW
from 6 kW) and three additional remote stations were added to the
facility. In this new configuration, the number of measured orbits has
increased from 2000-3000 per day (with the original CMOR) to $\sim5000$ per
day with CMOR II. Additionally, many orbits now have more than
three station detections, and so the accuracy of the overall orbital
measurements has improved.

Our meteor outburst survey follows the methodology previously employed
to detect longer-lived showers in CMOR orbital data by using a
3-dimensional wavelet transform (see \cite{browonwer10} for
details). All potential single day maxima detected by this approach in
CMOR data are correlated with known showers based on the original CMOR
shower catalogues \cite[]{browerwon08,browonwer10} and the working
list of meteor showers maintained by the International Astronomical
Union. We find that CMOR detects 1-2 outbursts each year, defined as
either unknown, intense, short-lived showers or significant
enhancements over normal activity from among previously known annual
streams.  Among the former category was the detection of the Daytime
Craterids in 2003 and 2008 \cite[]{wiebrower11} while the latter type
of outburst is well represented by the October Draconids in 2005
\cite[]{camvaubro06}. Full details of this complete outburst survey
will be published separately. Here we present details of the detection
of the Andromedids in 2011 as a separate study due to its unusual
nature and intensity.  In Table~\ref{ta:radar2011} we list the top
three strongest radiants detected by CMOR II using our wavelet
technique between solar longitudes $\lambda = 250-252\degree$ in 2011
(Dec 3-5). In all previous years either the NOO (November $\omega$
Orionids) or GEM (Geminids) were by far the dominant radiants during
this interval. The intensity of the shower is best represented by
$N\sigma$, the number of standard deviations the wavelet coefficient
is above the median background. Our normal single-day strength cutoff
establishing a probable shower maximum point is $8\sigma$. For
comparison, the Draconid outburst in 2005 produced an $N\sigma$ of 39
and the Daytime Craterid outbursts an $N\sigma$ of 36 and 33 in 2003
and 2008 respectively. During the 10 years of CMOR operation, other
than these showers, only the Andromedid return in 2011 has exceeded a
threshold of 30 $\sigma$ for a single day outburst. Note that in
Table~\ref{ta:radar2011} the strong radiant was not automatically
identified with the traditional Andromedid shower due to large
differences in the radiant; this disparity is explained in the
modelling section.

\figurenum{1}
\begin{figure}[htbp]
\plotone{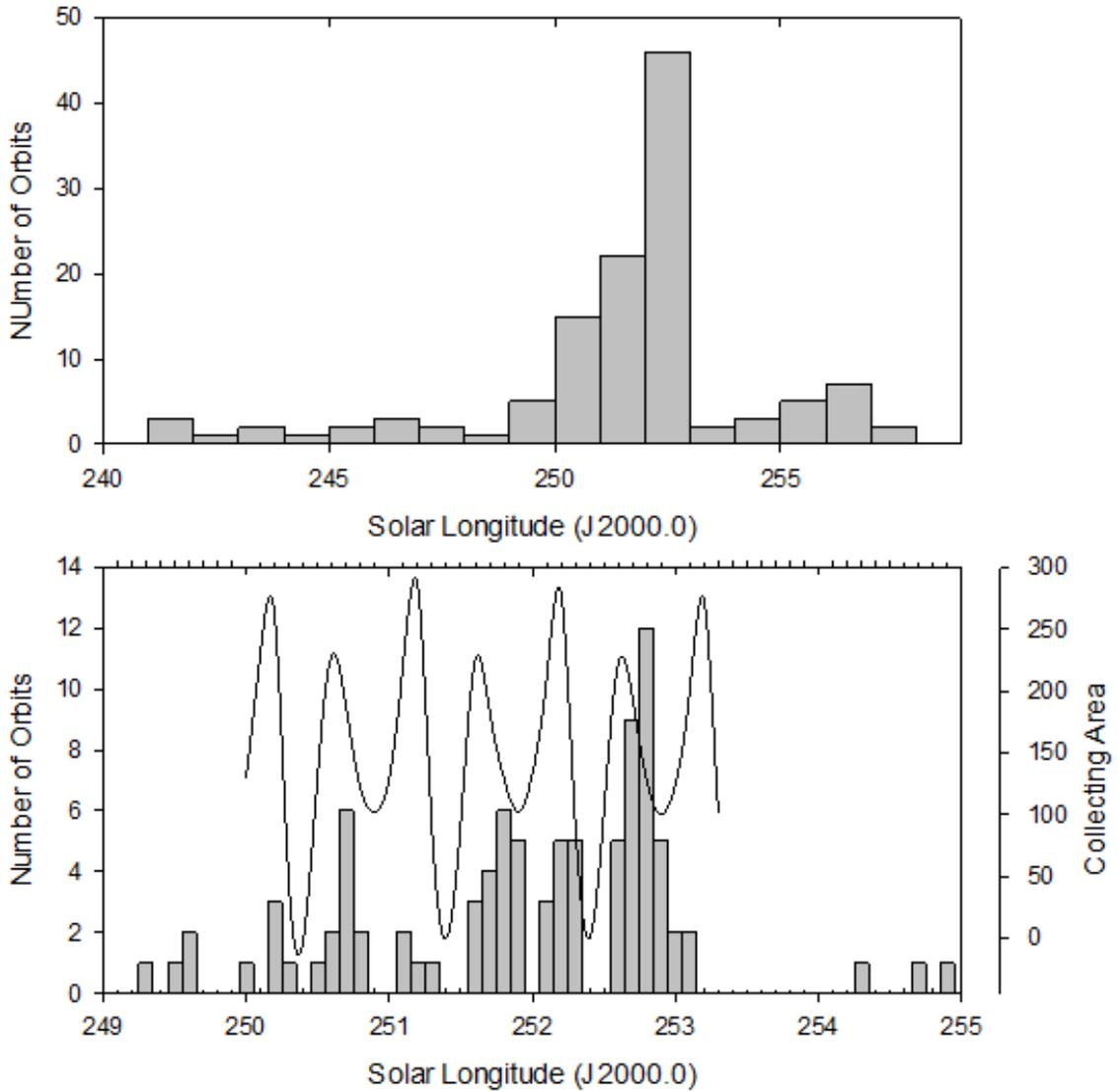}
\label{fi:observedsollon}
\caption{ CMOR's 2011 Andromedid meteor orbit count as a function of solar
  longitude. The bottom panel also displays the (time-varying)
  collecting area in square km. Note that there was a loss in
  transmitting power (and hence meteor counts) due to freezing rain
  beginning at $253.3\degree$ until $254.1\degree$.}
\end{figure}

\tablenum{1}
\begin{table}[p]
\centerline{
\begin{tabular}{lcccccc} 
\tableline \tableline
 $\lambda$ (J2000)   & RA ($\degree$)  & Dec ($\degree$)   & V$_g$ (km~s$^{-1}$) & $N\sigma$ & $N_{orb}$  & desig \\ \tableline \tableline
250 & 92.8  & +15.9 & 42.9 & 33.7 & 459 & NOO \\
    & 20.2  & +54.1 & 16.2 & 14.5 & 45  &     \\
    & 100.4 & +34.6 & 33.6 & 12.9 & 380 & GEM \\ \tableline
251 & 93.6  & +15.4 & 42.9 & 25.1 & 399 & NOO \\
    & 19.9  & +56.9 & 17   & 22.1 & 38  &     \\
    & 101.7 & +34   & 33.6 & 18.2 & 343 & GEM \\ \tableline
252 & 18.2  & +57.5 & 16.2 & 30.6 & 63  &     \\
    & 102.7 & +34.5 & 33.6 & 16.6 & 314 & GEM \\
    & 94.1  & +14.4 & 40.8 & 14.9 & 296 & NOO \\ \tableline
\label{ta:radar2011} \end{tabular}
}
\caption{The top three wavelet peaks in 2011 and associated showers
  from solar longitudes $\lambda$ from $250\degree-253\degree$. $V_g$
  is the geocentric velocity, $N\sigma$ represents
  the number of standard deviations of the wavelet detection above the
  usual background, and $N_{orb}$ is the number of orbits used to compute the wavelet coefficient. Two other showers are active at this time: GEM is the
  Geminids, NOO is the November $\omega$ Orionids: the lines without an
  entry in the designation column list the data for the 2011
  Andromedids outburst reported here. Radiants are geocentric J2000.}
\end{table}

\begin{figure}[htbp]
\plotone{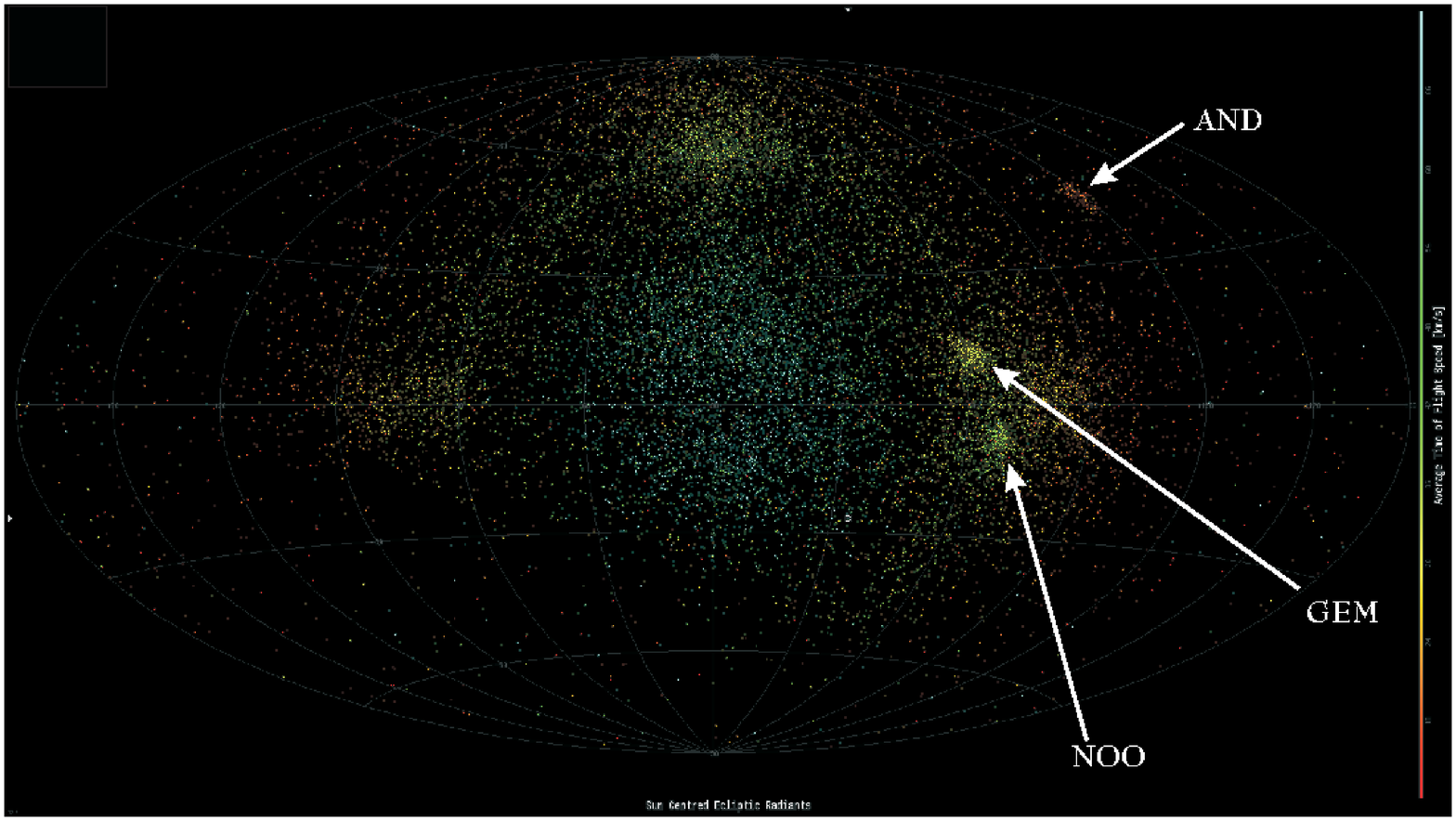}
\label{fi:observed-a}
\caption{CMOR radiants detected between solar longitudes of $250\degree-253\degree$, plotted in a Sun-centered ecliptic reference frame. The apex of the Earth's motion is at the centre of the plot. The color coding is apparent velocity.}
\end{figure}

\begin{figure}[htbp]
\epsscale{0.8}
\plotone{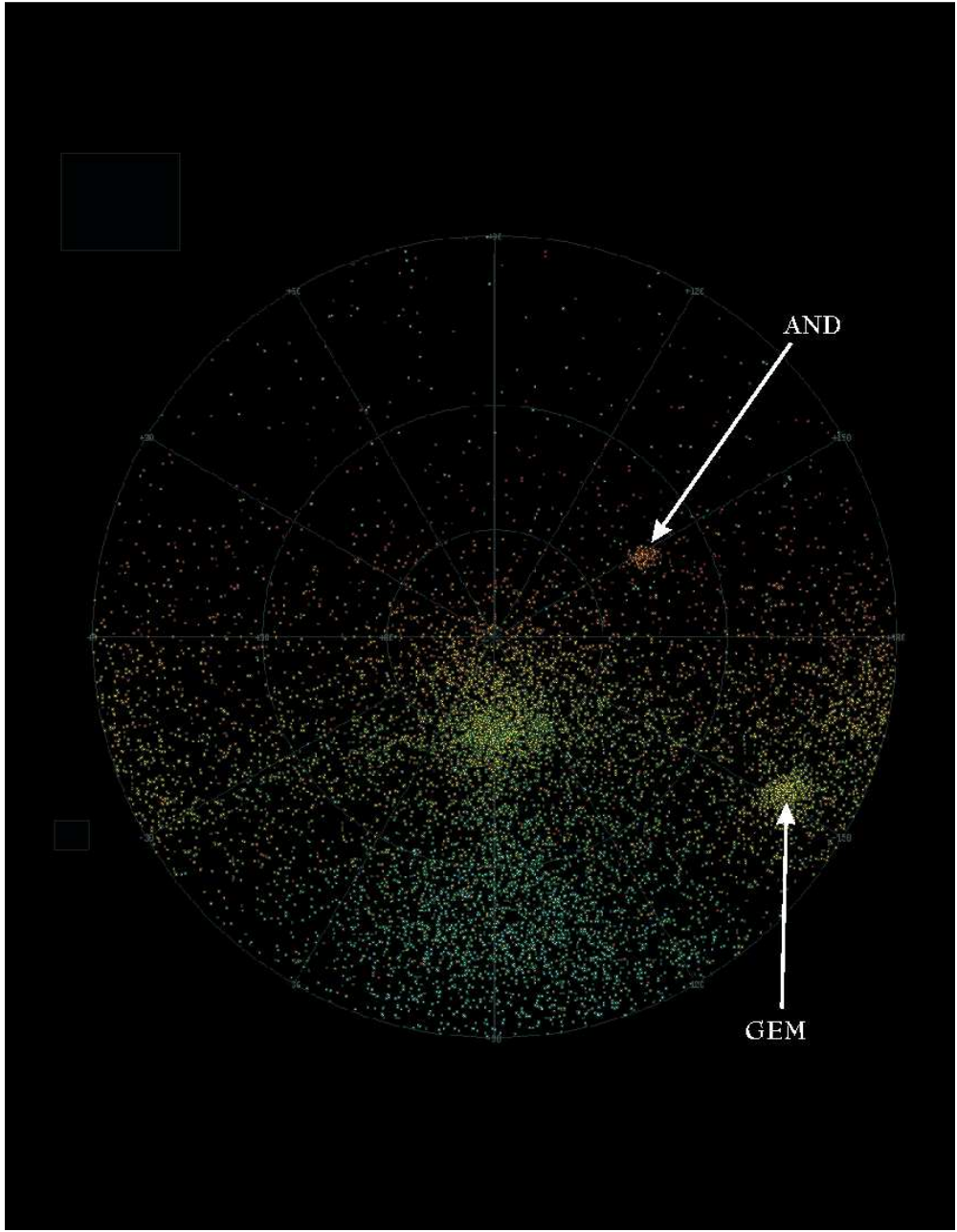}
\label{fi:observed-b}
\caption{CMOR radiants  detected between solar longitudes of $250\degree-253\degree$, plotted in a Sun-centered ecliptic reference frame, looking down on the north ecliptic pole (center).}
\end{figure}

The peak activity appears to have occurred between
$\lambda=252\degree-253\degree$, most likely between
$\lambda=252.7\degree- 252.8\degree$ (9-12 UT, Dec 5) based solely on
the number of orbits recorded from the shower
(Figure~\ref{fi:observedsollon}). The absolute peak flux at this time
was equivalent to a zenithal hourly rate (ZHR) of approximately 50.
The increase in numbers during this period is probably significant
given that the collecting area of the radiant at the time of maximum
as seen by CMOR II decreases by 40\% compared to the previous two hour
interval. However, whether the outburst continued through the next day
is uncertain as a period of freezing rain began near 0 UT Dec 6
($\lambda=253.2\degree$) and this led to the radar automatically
shutting off (due to high TX power reflection) until 18.5 UT Dec 6
(254.1$\degree$). Certainly the activity did not extend significantly
into the $\lambda=254\degree$ range, but we have almost no radar
observations (due to the weather and low radiant collecting area)
during $\lambda=253\degree- 254\degree$.  In total, 122 probable
Andromedid orbits were recorded during the interval from
$\lambda=240\degree-260\degree$, the majority (85) occurring between
$\lambda=250\degree-253\degree$. By way of comparison, between
$\lambda=250\degree-253\degree$ the average number of radiants within
10 degrees of this location having similar speeds between 2001 - 2010
was $\sim 10$ per year. Figure~\ref{fi:observed-a} shows all $\sim 14
000$ measured radiants in 2011 detected by CMOR II between
$\lambda=250\degree-253\degree$ in sun-centred ecliptic coordinates as
a Mercator projection; a polar projection is in
Figure~\ref{fi:observed-b}. The clump of Andromedid radiants is
obvious to the eye: it particularly stands out from the background
because of the very low radiant density at such a large elongation
from the Earth's apex. Also labelled are the Geminid and November
$\omega$ Orionid radiants. From the cumulative amplitude distribution
of the 122 echoes from the Andromedid radiant, we follow the basic
technique outlined in \cite{blacamwer11} to compute a shower mass
distribution index. The mass index $s$ is the power-law exponent in
the differential relation presumed to hold between the cumulative
number of meteoroids of mass $m$ or larger and their number such that
$dN=m^{-s}dm$ (see \cite{cep98} for a complete description). Note that
unlike the single-station selection technique used in
\cite{blacamwer11} which necessarily includes substantial
contamination from other radiants, we are able to select only those
radiants associated with the shower. We find $s=2.2$ for the stream,
substantially higher than found for other major streams with CMOR,
suggesting strongly that the outburst was rich in fainter meteors.

\figurenum{2a}

\figurenum{2b}

\figurenum{3}
\begin{figure}[htbp]
\plotone{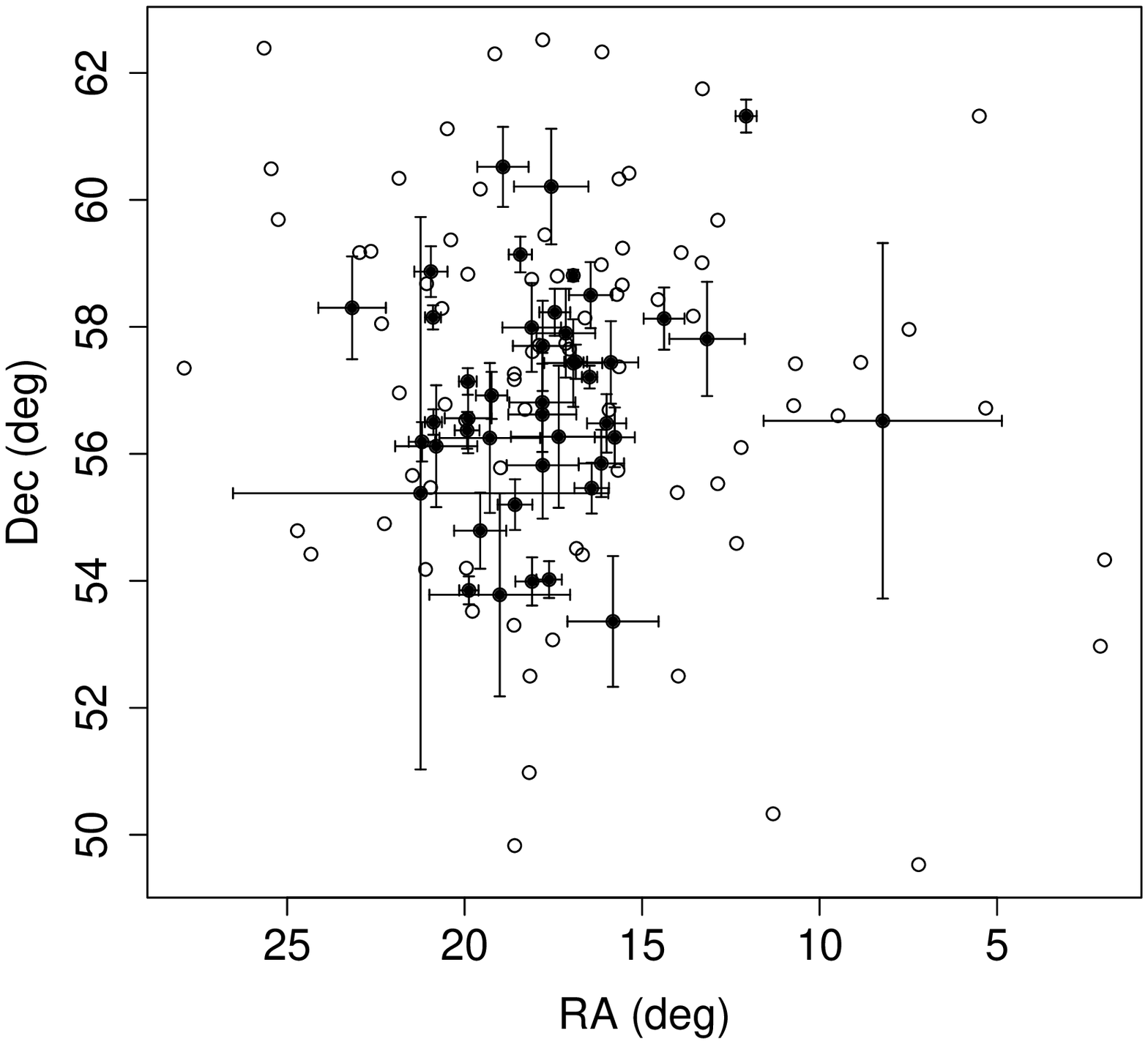}
\label{fi:observedradiants}
\caption{Geocentric right ascension and declination for all radiants (open circles) and for the highest quality radiants (black circles with errors) from CMOR. See the text for more details.}
\end{figure}

The radiant of all 122 probable Andromedids is shown in
Figure~\ref{fi:observedradiants}. Note that the errors per radiant are
found using a Monte Carlo technique which adds Gaussian noise to an
idealized model echo having the same set of signal-to-noise ratios,
decay times, heights and nominal time offsets at each outlying station
as the observed echo and then chooses the inflection time picks and
performs interferometry using the same algorithms used to process real
data, as described in \cite{werbro12}. The resulting errors represent
the standard deviation of the radiant position computed per orbit
among these simulations. This approach captures the large errors often
associated with poor geometry in the multi-station reflection process.
To better define the shower, we further select only those echoes
having errors in semi-major axis less than 20\% of the semi-major axis
value, according to the Monte Carlo process. This amounts to $\sim 50$
orbits total; of these, 90\% are in the $250\degree-253\degree$
interval. The radiant distribution for this high quality orbit group
is shown as well in Figure~\ref{fi:observedradiants}. It is clear that
there is a tight radiant near RA=18$\degree$ and Dec=$+56\degree$,
with a spread of order $5\degree$. Note that we have not corrected for
any radiant drift in this plot of observed geocentric radiant
positions; hence some of this spread is simply due to
drift. Interestingly, as part of the outburst survey, we also detected
a relatively strong single day maximum on November 27, 2008 associated
with the Andromedids (but at a lower declination than the 2011
outburst) - this is shown in Table~\ref{ta:radar2008}. The number of
radiants associated with this detection is small (30) and the shower
strength is lower than in 2011. This matches a weak shower seen in the
simulations, discussed in section~\ref{results}.

\tablenum{2}
\begin{table}[p]
\centerline{
\begin{tabular}{lcccccc} 
\tableline \tableline
            & $\lambda$ (J2000)   & RA ($\degree$)  & Dec ($\degree$)   & V$_g$ (km~s$^{-1}$) & $N\sigma$ & $N_{orb}$ \\ \tableline \tableline
CMOR        & 246   & 26.6 & +44.4 & 15.9 & 11.5 & 30 \\ \tableline
Simulation  & 251.6 & 28.8 & +47.4 & 16.0 & - & 77 \\ \tableline
\label{ta:radar2008} \end{tabular}
}
\caption{CMOR and simulation data for the 2008 Andromedid shower. See Table~\ref{ta:radar2011} for more details.} 
\end{table}

In addition to the CMOR observations, the Southern Ontario Meteor
Network \cite[]{browerkoh10} detected one possible member of the
outburst on Dec 3 at 04:24 UT of magnitude -1. Note that Dec 3 was the
only clear night in the interval 2011 Dec 3 - 8 in southern
Ontario. On the same night, a single station of the Canadian Automated
Meteor Observatory (CAMO) \cite[]{wercamwie12} recorded
3 possible members of the outburst between 0530 - 0730 UT Dec 3, all
fainter than +4. These optical observations are consistent with the
outburst being generally rich in faint meteors as suggested by the
steep mass index measured by the radar.

It is worth noting that the radiant of the 2011 shower reported here
is more typical of the `early' Andromedids of the 19th century, a
December shower whose radiant is near RA $=+20\degree$, Dec
$=+54\degree$ (actually in Cassiopeia) rather than the `modern'
Andromedids, a November shower whose radiant is near RA $=+25\degree$
Dec $=+44\degree$.   Precession of the meteoroid
stream has resulted in the displacement of the classical stream to its
current position.  Nonetheless, simulations (described later) reveal
that some slowly-precessing material released from 3D/Biela in the 17th
century is responsible for the outburst seen in 2011.

\section{Simulations} \label{simulations}

In order to better understand the nature of the 2011 outburst,
numerical simulations were performed of the parent comet. Simulations
of meteoroids released during each of Biela's perihelion passages up
to 200 years prior to its discovery were examined for clues as to the
origin of the material that produced the 2011 outburst. The properties
of the outburst in 2011 could be used to better refine earlier orbits
for 3D/Biela, though we are looking only for coarse agreement to
establish the broad strokes of the origin and evolution of the
streamlet seen in 2011.

The meteoroids were simulated within a Solar System of eight planets
whose initial positions and velocities were determined from the JPL
DE406 ephemeris \cite[]{sta98}. The particles were integrated with a
symplectic integration code \cite[]{wishol92} which handles close
encounters by the \cite{cha99} method. Some simulations were
duplicated with the RADAU method \cite[]{eve85} and the results were
found to be qualitatively the same. The speed of the symplectic code
allowed roughly ten times as many particles to be simulated, and these
`higher-resolution' results are reported here. The Earth-Moon are
simulated as a single particle at the Earth-Moon barycenter.  A time
step of 7 days is used in all cases.

The simulations were also run with a novel two-stage refinement
procedure. First we describe the initial stage, which parallels the
usual approach to such studies. The comet orbit is integrated
backwards to the desired starting point, 200 years prior to the
comet's discovery.  The comet is then integrated forward again,
releasing meteoroids at each perihelion passage as it does so. The
methodology of \cite{vaucoljor05} is followed whereby the size
distribution can be taken into account by applying a
statistical weight to the particles based on their mass index {\it after}
the simulations are completed.  In our simulations, at each
perihelion passage a number $M=1000$ particles is released in each of
the four size ranges of $10^{-5}-10^{-4}$~m, $10^{-4}-10^{-3}$~m,
$10^{-3}-10^{-2}$~m and $10^{-2}-10^{-1}$~m, extending from 10~microns
to 10 cm in diameter, so that the distribution of particle radii is
flat when binned logarithmically. The simulations include
post-Newtonian GR corrections and radiative ($i.e.$
Poynting-Robertson) effects. The ratio of radiative to gravitational
force $\beta$ is related to the particle radius $r$ (in $\mu$m)
through $\beta = 0.57/r$ following \cite{weijac93}, where we use a
particle mass density $\rho = 1000$~kg~m$^{-3}$.

The comet is considered active (that is, simulated meteoroids are
released) when at a heliocentric distance of 3~AU or less. While
active, particles are released from the parent comet in time from a
uniform random distribution, with velocities from the prescription of
\cite{crirod97}. Lacking specific information about the nucleus of
comet Biela (\eg \cite{tanferric00}), the Bond albedo of the comet
nucleus is taken to be 0.05, the nucleus and meteoroid densities
$1000$~kg~m$^{-3}$, the nucleus radius 1000~m, and the active fraction
of the comet's surface, 20\%.

The comet and all meteoroids are integrated until the simulation's end
point. The output is searched, and all meteoroids which pass
sufficiently close to the Earth's orbit during the period of time in
question are extracted and examined: this is our list of
``bulls-eyes''. This is the end of the first stage of the simulations
and to this point, the process follows the commonly-accepted procedure
for studying meteor showers.

The next stage refines the results by concentrating on those
meteoroids which are able to reach the Earth. Such methods have been
used before with great success. \cite{wuwil96}, \cite{ash99} and
\cite{mcnash99} used the fact that the orbit of comet 55P/Tempel-Tuttle
evolves only slowly to preferentially select Leonid shower meteoroids
that would impact the Earth for simulation, at a considerable saving in
computational time. Our method is similar but does not require the
parent orbit to be slowly evolving in time. Rather, in our refinement
stage, the list of ``bulls-eyes'' is used to populate a second set of
simulations. In this second set, the parent comet is integrated in
exactly the same manner as in the first stage, except that this time
meteoroids are {\it only} ejected near the initial conditions known to
produce bulls-eyes in the first simulation.

The details of the second stage procedure used are as follows. At each
time step during the second simulation, a check was made to see if a
``bulls-eye'' was produced in the same time step in the first simulation.
If so, $N$ new particles with orbits similar to the bulls-eye are
produced, where the calculation of $N$ is discussed below. All
particles have the same position (that of the nucleus, taken to be a
point particle). Each component of the meteoroid's velocity vector
relative to the nucleus is given a random kick of up to $\pm 10$\% of
that of the original bulls-eye, as is its $\beta$. These ``second
generation'' particles (and any others that are produced in later time
steps by other bulls-eyes) are then integrated forward in the usual
way. At the end of the simulation, those meteoroids which pass near
the Earth at the time under investigation are extracted. Invariably,
these contain far more meteoroids than the bulls-eye list of the
initial simulation.  As a result, a much clearer look at the regions
of phase space which produce the shower event in question is obtained
at relatively low computational cost.

The number $N$ of particles produced near a given bulls-eye is
calculated as follows. Let $M$ (always taken to be 1000 here) be the
number of particles released in a given perihelion passage in each of
the four size bins.  If there were $n$ bulls-eyes recorded in that
particular size bin during the current perihelion passage then each
is assigned a fraction $1/n$ of the $M$ particles assigned ($N=M/n$).
For example, if there are 10 bulls-eyes of sizes from $10^{-5}$ to
$10^{-4}$~m in the first simulation, then the region near each of
those 10 will be seeded with $M/10 =100$ particles in the
second-generation. If there had been 100 bulls-eyes in this perihelion
and size bin, then each bulls-eye would have been seeded with
$M/100=10$ particles in the second generation. Typically $\sim 1$\%
particles released in the first generation hit the Earth, and so each
bulls-eye is usually reseeded by $\sim100$ particles. 

This procedure is adopted, instead of say, simply replacing each
bulls-eye with a fixed number of particles because 1) it maintains a
constant computational load from simulation to simulation and 2) it
favours perihelion passages which produce few meteors: this hopefully
allows us to avoid missing possibly-rich perihelion passages that
might have been missed by the granularity of the first simulation.
The procedure adopted does make it slightly more complicated to
convert from the number of simulated meteors in a shower event to the
actual number, but this is easily accomplished by assigning a weight
to each simulated meteor that is proportional to $1/N$.

The effects of granularity in the first simulation are worth noting.
If a given perihelion passage produces no bulls-eyes in the initial
simulation, then at the refinement stage, no meteors at all will be
released during that passage. Thus the first simulation must sample
the available phase space well enough (\ie $M$ must be large enough)
or the refinement procedure will fail. It appears empirically that our
choice of $M=1000$ is sufficient to meet this condition in this case,
but the possibility that important regions of meteoroid ejection phase
space have been missed remains.

The list of bulls-eyes at the end of the first stage would ideally
consist entirely of particles which physically collide with the Earth;
however current computational limits prevent simulating the number of
particles needed to produce a statistically significant number of such
collisions. Thus we are forced to select our criteria more generously
and somewhat arbitrarily, though guided by experience. Here we have
required a bulls-eye to satisfy two criteria 1) the minimum orbital
intersection distance between the meteoroid's orbit and the Earth's
orbit should be less than 0.1~AU. Note that we do not use a nodal
distance but a true minimum in the inter-orbit distances. Though more
laborious to compute, it is more robust in the case of low-inclination
orbits where the node may be located a large distance from the closest
point of approach. 2) The meteoroid should be at its closest approach
to Earth within $\pm 30$~days of the shower date, here taken to be
December 4th 0h UT.  Thus the bulls-eye criteria specifically select
those meteoroids which are closest to Earth during the meteor shower
one is modelling, out of all the meteoroids simulated, of all
sizes, from all perihelion passages simulated.

An auxiliary simulation was performed in which the comet orbit was
integrated backwards for a thousand years in order to determine its
Lyapunov exponent using the algorithm of \cite{mikinn99}. This measure
of the chaotic time scale is necessary to understand over what
interval we can have confidence in our simulations.  The $e$-folding
time of Biela was found to be $\approx25$~years.  This short Lyapunov
time is typical of a Jupiter-family comet that has numerous close
approaches to that giant planet. We find that in our simulations
Biela's 1772 orbit, when integrated back 200 years, has close
encounters ($\le 5$~Hill radii) with Jupiter in 1747, 1711, 1664, 1652
and 1604. The last in 1604 is at $\approx 1.2$ Hill radii. Thus our
backwards integrations of 200 years go back eight Lyapunov times,
which stretches the limits of what we expect chaos to allow us to
calculate reliably, and so we do not attempt to go further back. Note
that the timing, particle population and radiant of the observed
trailet in 2011 may actually allow a better refinement of the early
orbit for 3D/Biela, though such a study is beyond the scope of the
current work. Using the timing and characteristics of contemporary
showers with known ejection age as a means to potentially constrain
the early orbit of a parent comet has been discussed before (\eg
\cite{vauwatsat11}).

The comet orbital elements used in these simulations are derived from
\cite{marwil08}. No non-gravitational forces due to outgassing (\eg
\cite{marsekyeo73}) were applied. Because the comet orbit evolved
significantly during the 80 years between when it was discovered
(1772) and the last observation of the fragments (1852) we did not use
a single set of orbital elements for the comet. Rather, the 1772
discovery orbit was used as the starting point for all perihelion
passages prior to this time back 200 years to 1572. The perihelion
passages from 1772 to 1852 were modelled in parallel three times,
using the orbit of 1832, as well as the final orbits of fragments A
and B from 1852, their last apparition. These are all listed in
Table~\ref{ta:elements}. The multiple models for the end of comet
Biela's lifetime allow a careful exploration of the phase space since
the dynamical effects of fragmenting and additional outgassing, not to
mention the ever-present perturbations due to Jupiter, make the
simulation of this very interesting part of Biela's life difficult. We
will see that the different fragment orbits produce different results,
as will be discussed in more detail in section~\ref{results}.

\tablenum{3}
\begin{table}[p]
\centerline{
\begin{tabular}{llccccccc} 
\tableline \tableline
n & Name        & Peri date & $q$ (AU) & $e$ & pd (yr) & $\omega$ ($\degree$) & $\Omega$ ($\degree$) & $i$ ($\degree$) \\ \tableline \tableline
1 & 3D/1772 E1  & 1772 Feb. 17.675  & 0.99038 &  0.72588 &  6.87&  213.340  & 260.942 &   17.054 \\
2 & 3D/1832 S1  & 1832 Nov. 26.6152T& 0.879073&  0.751299&  6.65&  221.6588 & 250.6690&   13.2164\\
3 & 3D-A        & 1852 Sept. 23.5432T& 0.860594&  0.755828&  6.62&  223.1890 & 248.0070&   12.5488 \\
4 & 3D-B        & 1852 Sept. 24.2212T& 0.860625&  0.755879&  6.62&  223.1912 & 248.0043&   12.5500 \\ \tableline
\label{ta:elements} \end{tabular}
}
\caption{A list of the orbital elements used for comet Biela and its fragments in this paper. From \cite{marwil08}.}
\end{table}

\section{Simulation results and comparison with observations} \label{results}

\begin{figure}[htbp]
\plotone{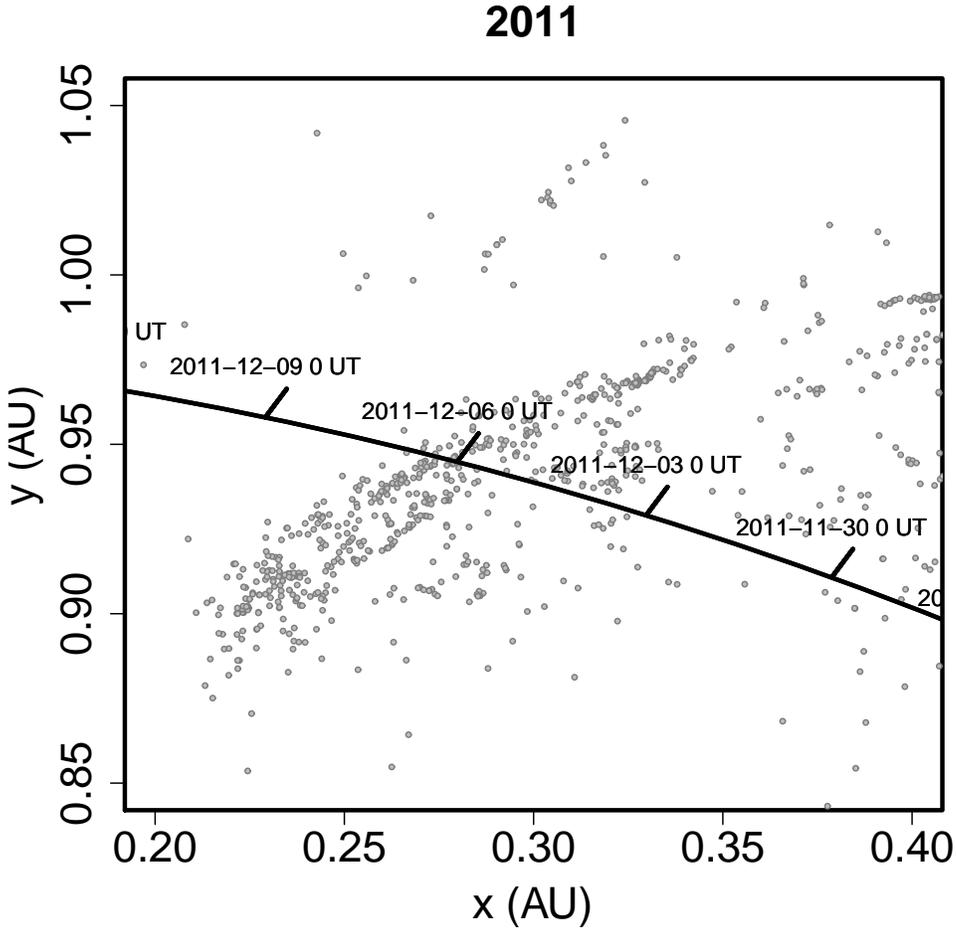}
\label{fi:footprint}
\caption{The nodal intersection points of simulated meteoroids arriving at the Earth in 2011. The Earth's orbit
is the heavy black curve with the time of the planet's passage labelled.}
\end{figure}

\begin{figure}[htbp]
\plotone{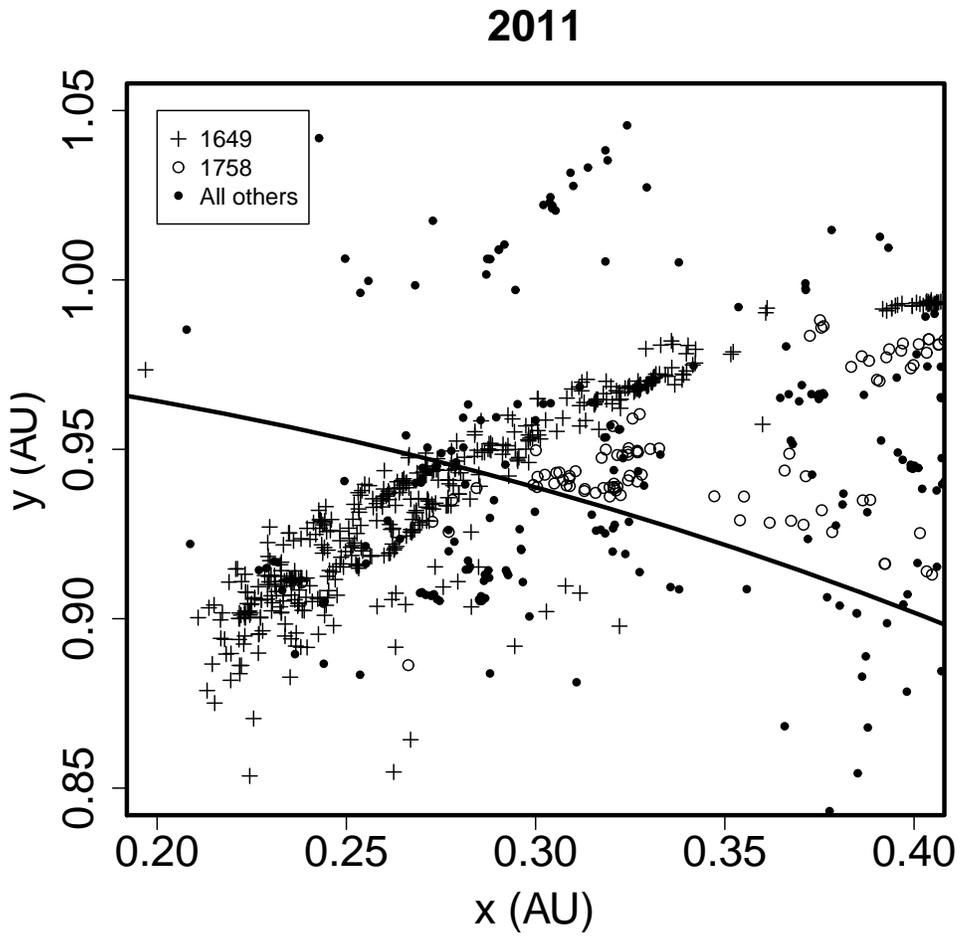}
\label{fi:footprintb}
\caption{The nodal intersection points of simulated meteoroids arriving at the Earth in 2011, identified by the
perihelion passage when they were produced.}
\end{figure}

\begin{figure}[htbp]
\plotone{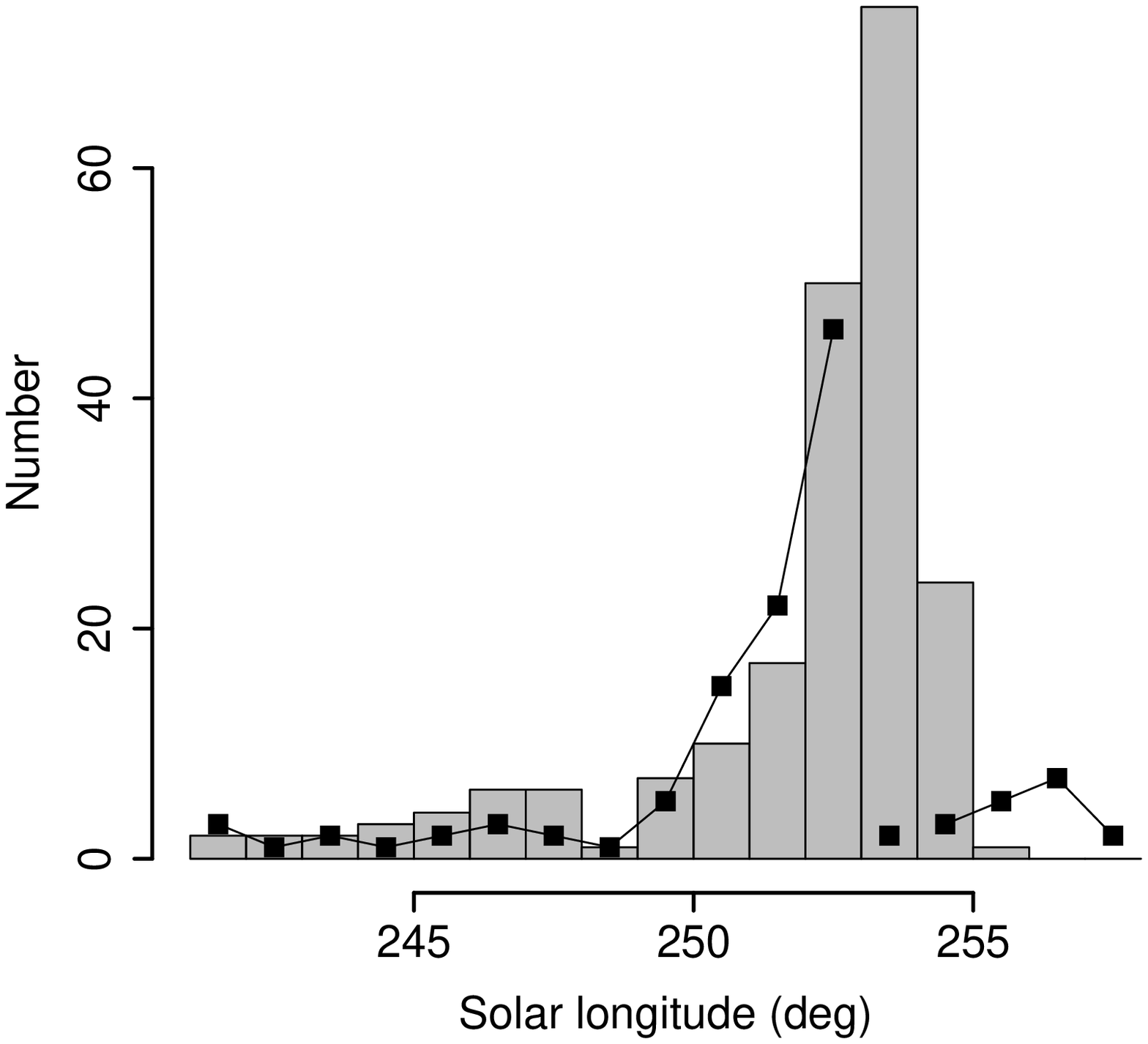}
\label{fi:arrivaltimes}
\caption{Counts of Andromedids meteors seen by CMOR (black squares) together with those of simulated meteoroids passing no more than 0.01~AU from Earth's orbit (grey histogram). A gap in the black line connecting the CMOR data at solar longitudes 253-254$\degree$ indicates the loss of sensitivity due to antenna icing.}
\end{figure}

The simulation output for the year 2011 are presented in
Fig.~\ref{fi:footprint}, which shows the nodal intersection points of
the simulated meteoroids overplotted on the Earth's orbit. A band of
meteoroids from the 1649 perihelion passage (as determined from the
backward integration of the 1772 orbit of 3D/Biela) sits astride the
Earth's orbit, along with a smaller grouping from the 1758 passage
(see Fig.~\ref{fi:footprintb}). The timing of the arrival of these
particles matches well with the CMOR observations
(Fig.~\ref{fi:arrivaltimes}) particularly when the loss of radar
sensitivity due to transmitter icing near $\lambda \approx
253-254\degree$ is considered. As a result of this match, we conclude
that the 2011 appearance of the Andromedids resulted from dust
produced by the 1649 perihelion passage of the parent comet, and other
details (discussed below) support this result.

\figurenum{4a}

\figurenum{4b}

\figurenum{5}

\figurenum{6}
\begin{figure}[htbp]
\plotone{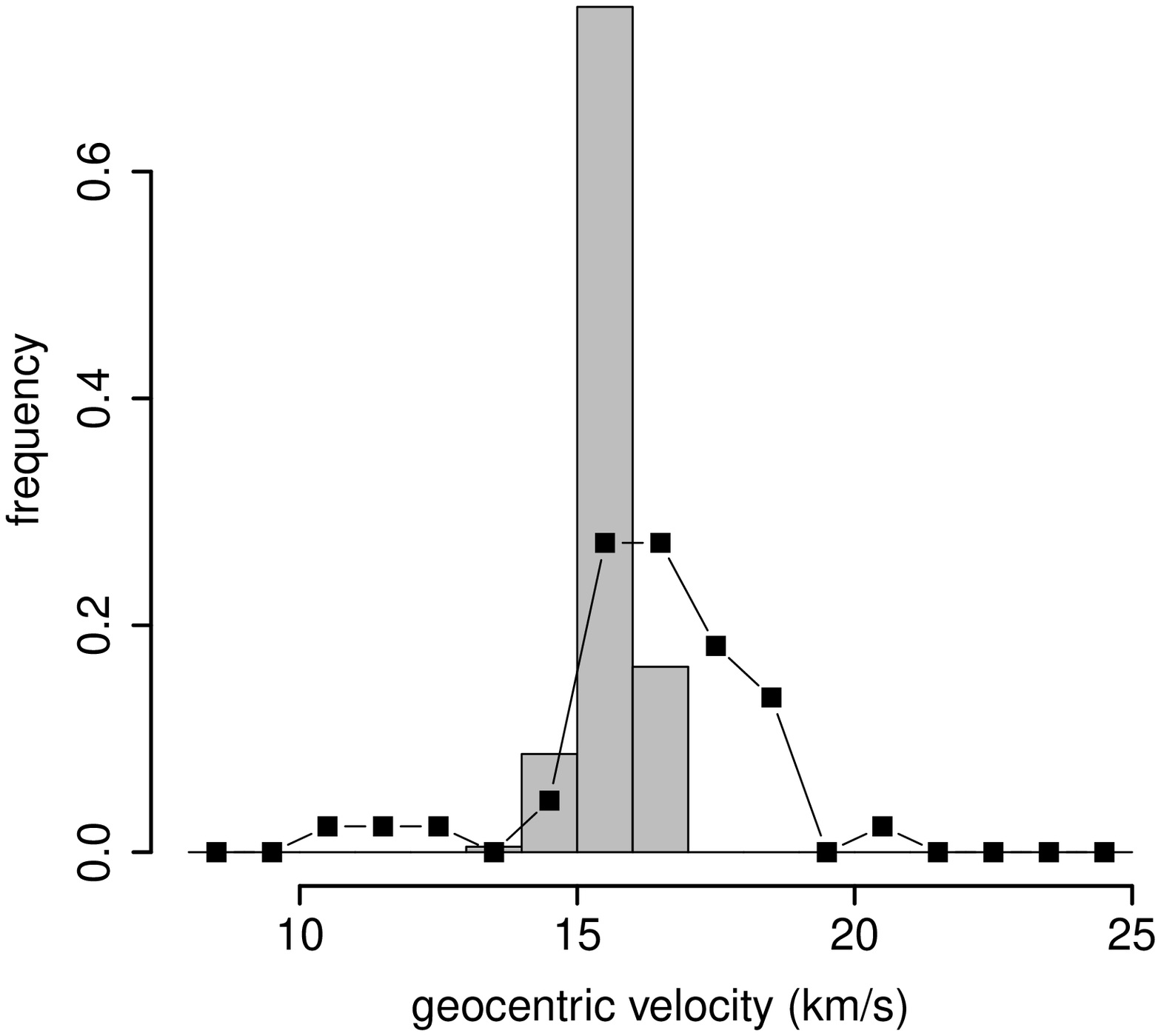}
\label{fi:vel}
\caption{The measured geocentric velocities of the 44 highest quality Andromedids measured by CMOR (black squares) and the simulated 2011 Andromedids (grey histogram).}
\end{figure}

The velocity of the CMOR-observed meteors is shown in
Figure~\ref{fi:vel}, as are those of the simulations. They are
consistent with each other, both concentrated near 16~km~s$^{-1}$ though
there is more spread in the radar data as would be expected from
measurement errors.

The size distribution is also qualitatively consistent with CMOR
observations. The simulated size distribution at Earth is concentrated
at small sizes, with no particles larger than 100~$\mu$m though our
simulations include particle with radii up to 10~cm. CMOR saw
particles which were somewhat larger than this: the typical sizes for
CMOR-detected Andromedids is 500~$\mu$m at an assumed density of
1000~kg/m$^3$ but the steep measured mass index of $s=2.2$ (see
Section~\ref{observations}) is consistent with a shower rich in small
meteors over larger ones.

\begin{figure}[htbp]
\plotone{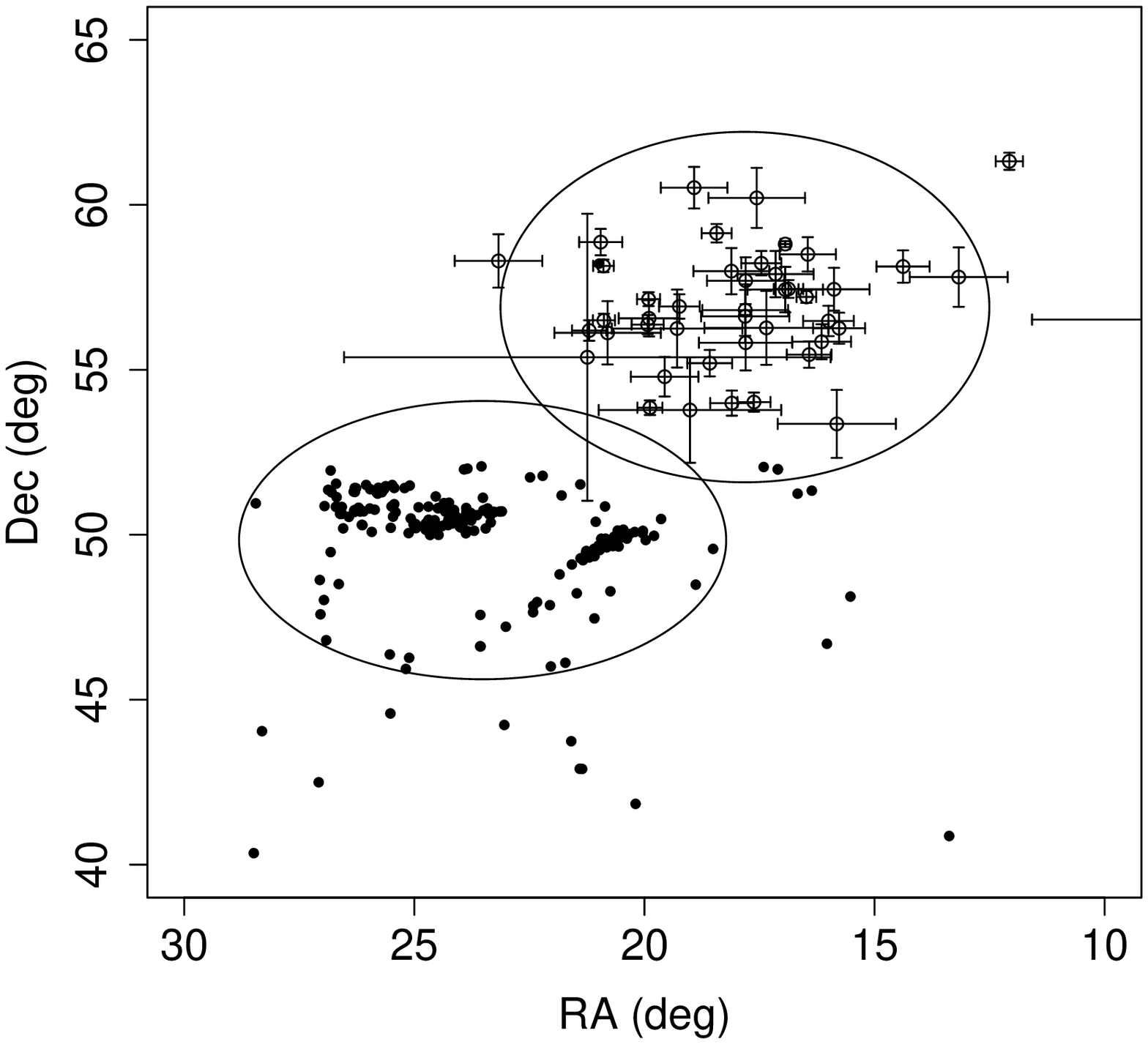}
\label{fi:radiants}
\caption{The right ascension and declination of the highest quality radiants from CMOR (open circles with error bars) along with the simulated radiants for the 2011 Andromedids (black circles). The ellipses show two standard deviations in RA and Dec for both radiants.}
\end{figure}

The radiants are shown in Fig.~\ref{fi:radiants}. The locations of the
simulated and observed radiants differ by about $\sim 8$ degrees. This
likely reflects the remaining uncertainty in 3D/Biela's orbit in 1649
compared to our adopted orbit as well as uncertainty in our
deceleration correction for the observed meteors between observed
in-atmosphere and estimated out-of-atmosphere speeds
\citep{brojonwer04}. In the simulations, two different radiants
separated by a few degrees are seen, one originating from the 1649
perihelion passage, the other from 1758.  The mean orbits of the two
simulated radiants are shown in Table~\ref{ta:radiants}: the numbers
are similar though not identical.

\figurenum{7}

\tablenum{4}
\begin{table}[p]
\centerline{
\begin{tabular}{cccccccccc} 
\tableline \tableline
Name & RA ($\degree$) & Dec ($\degree$) & $a$ (AU) & $a_{\beta=0}$ (AU) & $q$ (AU) & $e$ & $i$ ($\degree$) & $\Omega$ ($\degree$) & $\omega$ ($\degree$) \\ \tableline \tableline
CMOR     &  18.2 & 57.4  & 3.78 & 3.81 & 0.902 & 0.76  & 18.3 & 253.5  & 216.3 \\
$\pm$    &   2.6 &  2.2  & 0.71 & 0.71 & 0.012 & 0.04  & 1.0  &  2.4   & 3.1   \\ \hline
Sim: all &  23.6 & +50.2 & 3.78 & 3.40 & 0.894 & 0.763 & 14.5 & 252.5 & 218.1 \\
$\pm$    &  2.5  & 1.3   & 0.12 & 0.17 & 0.011 & 0.007 & 0.4  & 2.1    & 1.9   \\ 
Sim: 1649& 24.9  & +50.7 & 3.70 & 3.44 & 0.891 & 0.759 & 14.7 & 253.7 & 218.5 \\ 
$\pm$    &  1.1  & 0.4   & 0.02 & 0.18 & 0.007 & 0.001 & 0.2  & 0.6    & 0.8   \\ 
Sim: 1758&  21.2 & +49.2 & 3.91 & 3.31 & 0.905 & 0.769 & 14.1 & 250.7 & 217.2 \\ 
$\pm$    &  0.9  & 1.2   & 0.04 & 0.08 & 0.007 & 0.003 & 0.2  & 1.8    & 1.8   \\ \tableline
\label{ta:radiants} \end{tabular}
}
\caption{Numerical values for the radiants shown in   Fig.~\ref{fi:radiants}. The line beginning with $\pm$ indicates one
  standard deviation in the quantities above it. Note that the  orbital elements take into account the effect of
  radiation pressure which is non-negligible for some particles ( $\beta \las 10^{-2}$). The orbital element
most sensitive to $\beta$ is the semimajor axis $a$: its value under the assumption of $\beta=0$ is shown in the
  column labelled $a_{\beta=0}$ for comparison }
\end{table}

\figurenum{8}
\begin{figure}[htbp]
\plotone{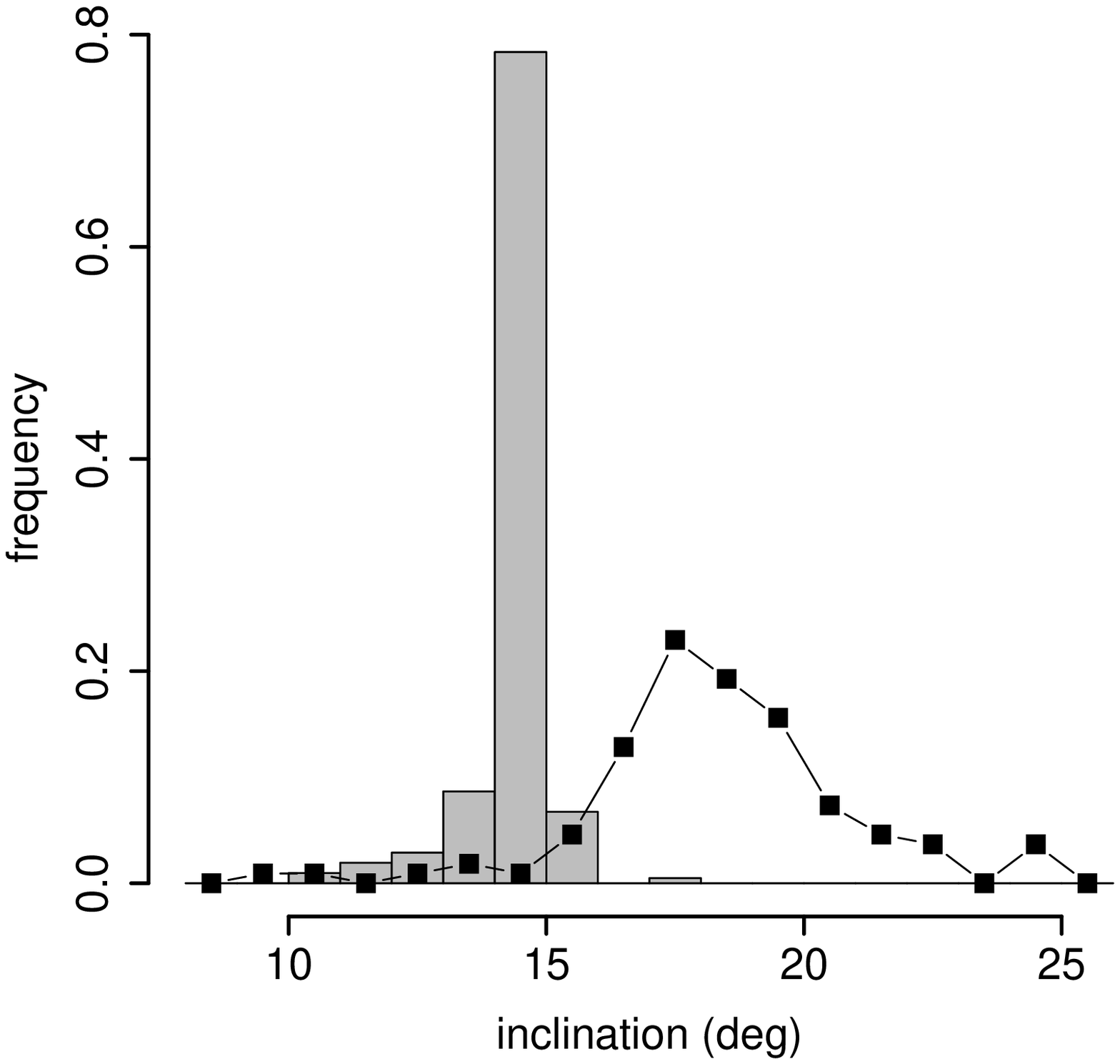}
\label{fi:inclination}
\caption{The CMOR-measured (black squares) inclinations along with those of the simulated (grey histogram) Andromedids.}
\end{figure}

The inclination distribution is shown in Fig.~\ref{fi:inclination};
along with the observed distribution. The observed is slightly higher
in inclination on average (by about 3$\degree$), but again, such small
deviations may reflects the remaining uncertainty in 3D/Biela's orbit
in 1649 compared to our adopted orbit compounded by our uncertainty in 
deceleration correction.

Given the nodal footprint of the 1649 perihelion passage, the
similarity of the times of arrival to observations, and the match
between the radar derived sizes, radiants and inclinations and those
of the simulations, we conclude that the 2011 Andromedids shower can
be traced primarily to the release of material by comet Biela during
its 1649 perihelion passage and that the shower is due to small
particles released during that passage having unusually favourable
dynamical delivery efficiencies to the Earth in 2011.

The meteoroids which reach the Earth from the 1649 perihelion passage
come primarily from the pre-perihelion leg rather than post-perihelion
(3:1 ratio), though they are otherwise distributed throughout the
comet's active phase ($r < 3$~AU).

The 2011 shower is more like the early appearances of the Andromedids
prior to the mid-19th century and this is tied to the relatively slow
precession rate of this particular streamlet. In the simulations, the
material in the 2011 appearance precesses less than that of the Biela
dust complex as a whole. Though we have not investigated the process
in detail, we do find that the portion of the dust stream involved is
trapped in a 3:5 mean motion resonance with Jupiter which likely
accounts for its differential evolution.

Since the 1649 perihelion passage seems to be the dominant source of
the meteoroids observed during the 2011 Andromedids shower, we
examined the literature for any evidence of particularly strong dust
production by comet Biela at this time. Of course, Biela was not
discovered until 1772, so no observations that can be definitively
linked to this object exist. We also checked the extensive compilation
of historical comet observations of \cite{kro99}, but no comets of any
kind are listed as having been seen between 1648 to 1651 inclusive.

\figurenum{9}
\begin{figure}[htbp]
\plotone{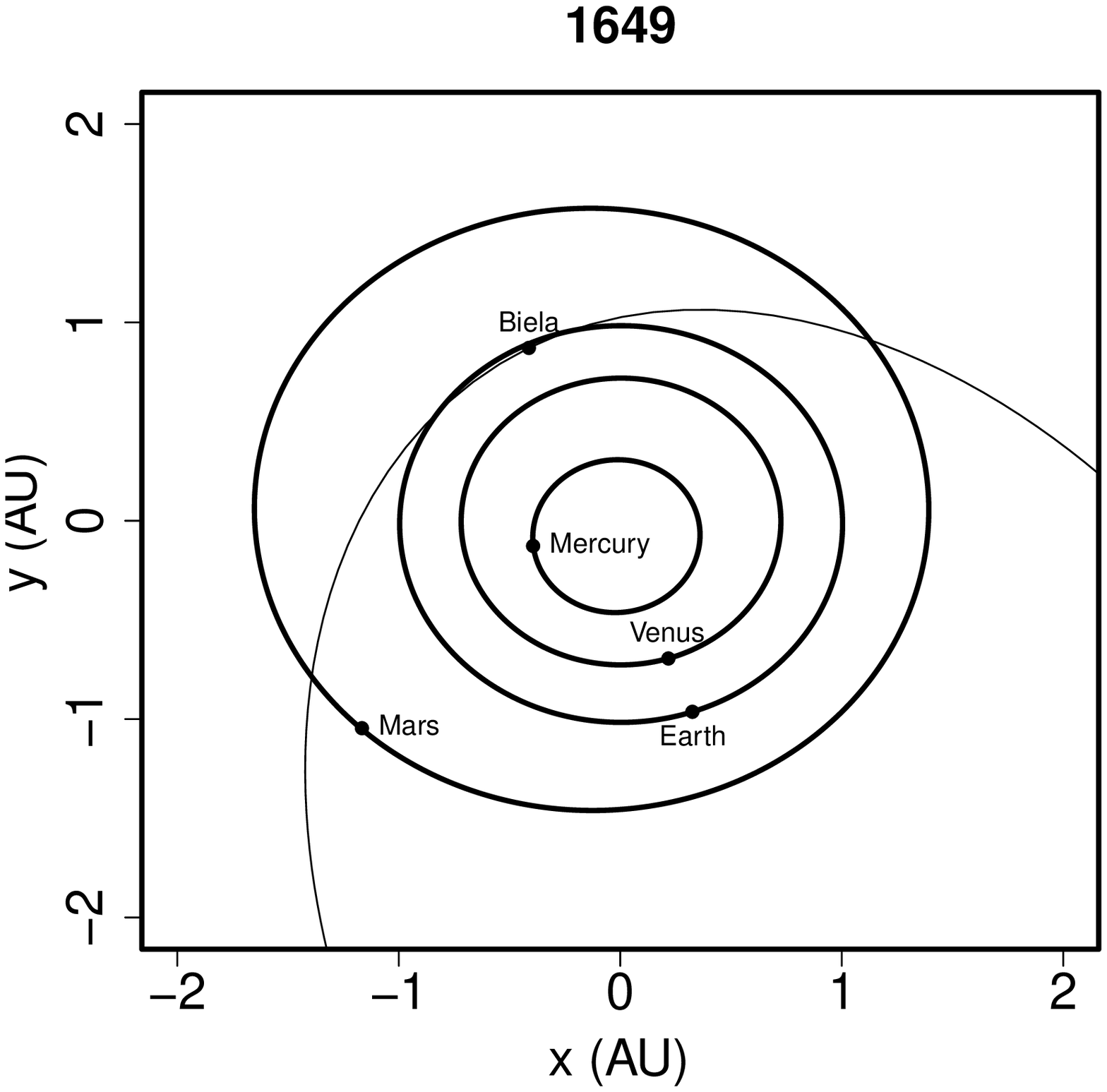}
\label{fi:1649circumstances}
\caption{The positions of the planets and comet Biela as the latter passed perihelion in 1649.}
\end{figure}

In our simulations, Biela passes perihelion in 1649 near July 5 (with
a $\pm7$~day uncertainty owing to our choice of time step), at which
time it would have been behind the Sun as seen from the Earth
(Fig.~\ref{fi:1649circumstances}). Thus a particularly active
apparition of the comet or even a fragmentation event in that year is
not impossible despite the fact that it passed unobserved. However, a
very high dynamical transfer efficiency for small meteoroids released
in 1649, coupled with proximity to the 3:5 MMR with Jupiter may be the
simplest explanation.

\tablenum{5}
\begin{table}[p]
\centerline{
\begin{tabular}{lccccccc} 
\tableline \tableline
 year & peri & RA ($\degree$) & Dec ($\degree$) & $\tau$ & $\lambda$ ($\degree$) & parent & $\langle \log_{10}(r) \rangle$ \\ \tableline
\bf 2001 & \bf 1649 & \bf 29 & \bf +42 & \bf 0.28 & \bf 249 & \bf  1772 & \bf -4.0$\pm$0.2\\ 
2002 & 1852 & 27 & +36 & 4.4 & 238 &  1852-A \& B & -4.2$\pm$0.2\\ 
\bf 2004 & \bf 1649 & \bf 25 & \bf +52 & \bf 1.8 & \bf 255 & \bf  1772 & \bf -4.1$\pm$0.2\\ 
2005 & 1812 & 27 & +40 & 5.8 & 245 &  1832, 1852-A \& B & -4.0$\pm$0.2\\ 
\bf 2008 & \bf 1649 & \bf 29 & \bf +47 & \bf 0.66 & \bf 252 & \bf  1772 & \bf -3.9$\pm$0.2\\ 
2010 & 1689 & 25 & +50 & 3.5 & 254 &  1772 & -4.1$\pm$0.1\\ 
2010 & 1846 & 27 & +37 & 6 & 241 &  1852-A \& B & -4.1$\pm$0.2\\ 
\bf 2011 & \bf 1649 & \bf 25 & \bf +51 & \bf 1 & \bf 253 & \bf  1772 & \bf -4.2$\pm$0.3\\ 
2012 & 1812 & 27 & +38 & 4 & 244 &  1832, 1852-A \& B & -4.0$\pm$0.2\\ 
2016 & 1852 & 27 & +36 & 3.5 & 239 &  1852-A \& B & -4.0$\pm$0.09\\ 
2018 & 1852 & 27 & +36 & 3.3 & 239 &  1852-A \& B & -4.0$\pm$0.1\\ 
\bf 2018 & \bf 1649 & \bf 24 & \bf +50 & \bf 0.71 & \bf 254 & \bf  1772 & \bf -4.1$\pm$0.1\\ 
2019 & 1819 & 26 & +37 & 12 & 241 &  1832, 1852-A \& B & -4.1$\pm$0.1\\ 
2022 & 1656 & 30 & +46 & 4.8 & 249 &  1772 & -3.7$\pm$0.2\\ 
\bf 2023 & \bf 1649 & \bf 29 & \bf +47 & \bf 4.1 & \bf 250 & \bf  1772 & \bf -3.8$\pm$0.2\\ 
\bf 2027 & \bf 1649 & \bf 25 & \bf +51 & \bf 0.36 & \bf 254 & \bf  1772 & \bf -4.2$\pm$0.2\\ 
\bf 2034 & \bf 1649 & \bf 25 & \bf +50 & \bf 0.28 & \bf 254 & \bf  1772 & \bf -4.2$\pm$0.2\\ 
2035 & 1656 & 29 & +44 & 4.1 & 247 &  1772 & -3.7$\pm$0.3\\ 
\bf 2036 & \bf 1649 & \bf 29 & \bf +45 & \bf 3.6 & \bf 248 & \bf  1772 & \bf -3.9$\pm$0.2\\ 
\bf 2041 & \bf 1649 & \bf 25 & \bf +49 & \bf 0.27 & \bf 254 & \bf  1772 & \bf -4.1$\pm$0.2\\ 
2043 & 1636 & 29 & +44 & 2.8 & 247 &  1772 & -3.6$\pm$0.2\\ 
2045 & 1852 & 27 & +35 & 5.1 & 238 &  1852-A \& B & -4.1$\pm$0.2\\
\label{ta:simshowers} \end{tabular}
}
\caption{Possible appearances of the Andromedids showers in the time
  interval 2000-2047. Those associated with dust produced by the 1649
  perihelion passage of 3D/Biela are indicated in bold. The column
  labelled 'parent' indicates the parent orbit of the shower (listed
  in Table~\ref{ta:elements}). The mean of the $\log_{10}$ of the
  radius (in meters) of the particles is also shown. See the text for
  more details.}
\end{table}

If the strong activity in 2011 was indeed produced by the 1649
perihelion passage, then one might expect other years where debris
from this particular perihelion passage is close to Earth would
produce similar showers.  To this end, we examined the simulations in
the years 2000 - 2047. The results are listed in
Table~\ref{ta:simshowers} which includes the shower details as well
the mean and standard deviation of the $\log_{10}$ of the radius in
meters of the particles comprising the shower. This quantity is not
weighted by the mass index but is calculated from the original
distribution of particles (which is flat in logarithmic space, see
Section~\ref{simulations}) and so provides an reasonable measure of
the typical particle sizes. Table~\ref{ta:simshowers} also lists a
transfer efficiency $\tau$. This quantity simply sums the weights
assigned to the simulated meteors (normalized to 1 for the 2011
shower) to provide an estimate of the efficiency with which dust is
transferred from the comet to the Earth during any given perihelion
passage. A shower with a large transfer efficiency does not
necessarily translate into a strong shower at Earth as there remains
the unknown factor of the cometary dust production: a large $\tau$
will not avail if the comet produces little dust.

Table~\ref{ta:simshowers} also lists the parent object, that is, which
of the four simulated parent orbits (listed in
Table~\ref{ta:elements}) produced the shower. Each parent orbit
contributes to different radiants in different years, a testimony to
the heavily-perturbed environment in which the meteoroid streams from
3D/Biela exist. The 1772 orbit of Biela produces showers at times
quite different from those of the 1832, 1852-A and 1852-B orbits; thus
future studies of the Andromedids complex will require careful
attention to the evolving orbit of the parent.

Though the comet dust production is expected to vary from perihelion
to perihelion, showers resulting from the 1649 perihelion passage are
all produced from the same dust release. Thus we expect that our
calculated transfer efficiency $\tau$ provides a reasonable estimate
of activity of these showers relative to the 2011 appearance. Our
simulations indicate weak to moderate activity ($\tau < 1$) in 2001,
2008, 2018, 2027, 2034 and 2041 as well as moderate to strong activity
($\tau >1$) in 2004, 2011, 2023 and 2036. The fact that the
simulations reproduce the correct activity ratio for both the 2008 and
2011 showers (the observed ratio of 2008 to 2011 ZHRs $\approx 30/50 =
0.6$ while the simulated $\tau_{2008} \approx 0.7$) give us some
confidence in the strength predictions for dust arising from Biela's
1649 perihelion passage at least, though 2018 will provide the first
opportunity to confirm these predictions after the fact. The future
showers in 2023 and 2036 are both four times stronger in our
simulation than that of 2011 and observers should be alert to these
appearances.

Of the predicted simulated showers, CMOR detected the 2008 appearance
of the shower (as was mentioned in section~\ref{observations}). The
simulation prediction of the 2008 shower was late ($\lambda \approx
252$ versus the observed peak at $\lambda \approx 246$) while the 2011
prediction was very close in time to that observed.  Simulations were
closer to the observed radiant in 2008 than in 2011 (with an angular
separation of $3\degree$ from the center of the observed radiant
versus $8\degree$ in 2011). Both simulations and radar results in 2008
show a radiant much closer to the `current' Andromedids radiant.
Simulations reveal that the 2008 outburst was comprised largely of
meteoroids released during the 1649 perihelion passage, reinforcing
our conclusion that this perihelion passage is an important
contributor to the Andromedids at Earth.

We also searched our simulations for evidence of showers other
perihelion passages of comet Biela. The strongest of these are also
listed in Table~\ref{ta:simshowers}. The CMOR database was checked at
all these dates $\pm$ 10 days from 2002 onwards: only 2008 activity
shows up.  The earlier years (2002-2004) had fewer orbits than later
years so the statistics are not as good - a weak shower could easily
have been missed in those years. In particular, the CMOR UHF links
were heavily attenuated in 2004 at the predicted time of the peak, so
any shower in the day or two around this period would be very hard to
detect from orbit data alone unless it was much stronger than 2011.
However, we would expect CMOR to have seen the apparitions in 2005
onwards if they did in fact occur.

Some of the simulated showers have $\tau$ larger than one and yet were
not observed. This may be telling us that these perihelion passages
did not produce much dust. Interestingly, the simulations show high
dust transfer efficiencies from Biela's fragments (1852-A, 1852-B) to
Earth in some years (\eg in 2010 from the 1846 passage of fragments A
and B), but these showers were not observed. Since the fragmentation
event was associated with strong Andromedid showers in the late 1800s,
one would expect that dust production was high at these times, and so
low dust production seems an unlikely explanation for the absence of
these appearances. Our adopted model for dust ejection is not
specifically tailored to break-up and that may play a role.  Also,
more active non-gravitational forces present after fragmentation may
have perturbed the fragment orbits in ways not modelled here.  We do
note that all the shower appearances listed in
Table~\ref{ta:simshowers} are rich in small meteoroids, typically
$\sim 100\mu$m, even smaller than those seen by CMOR in 2011. Thus it
appears that Earth-intersecting dust from 3D/Biela may be doubly
difficult to see because of small particle sizes and low
($\sim16$~km~s$^{-1}$) geocentric velocity.

Though the simulations show possible activity by perihelion passages
other than that of 1649, it is much more difficult to extrapolate this
to a prediction of real meteor activity. We have no evidence that
these other perihelion passages released dust in quantities comparable
to that of the 1649 passage.  So we consider predictions of Andromedid
showers resulting from the 1649 passage to be more robust than those
originating from other perihelia of the parent comet, as at least two
of the predicted outbursts from 1649 have definitely been detected in
recent years. Nonetheless, we have listed the strongest of these in
Table~\ref{ta:simshowers} as their observation (or not) at Earth will provide
useful insight into this comet's dust production.

\section{Conclusions} \label{conclusions}

The Canadian Meteor Orbit Radar (CMOR) detected an appearance of the
Andromedids meteor shower in early Dec 2011 at a radiant position
resembling that of the `classical' Andromedids radiant of the
1800's. The radiant was at RA=$18.2\degree$, Dec$=+57.5\degree$ at its
observed peak at a solar longitude near $\lambda=252.8\degree$ (2011
Dec 5 12 UT). A total of 122 meteors were observed and the associated
peak ZHR$\approx$50; predominantly small particles ($\sim 500\mu$m)
were seen.  A weaker shower appearance (ZHR$\approx30$) in 2008 on a
different but nearby radiant was subsequently found in the CMOR
outburst data base.

Simulations indicate that these showers arose from the 1649
perihelion passage of comet 3D/Biela, and the simulation's timing,
radiant and relative strengths all coarsely match those observed by
CMOR. The meteoroids producing the 2011 shower were trapped in 3:5
resonance with Jupiter which may account for their slower precession
and the radiant location closer to that of early appearances of the
shower.

Other appearances of the Andromedids are forecast, the next
originating from the same perihelion passage of the parent as that
which produced the 2011 shower will occur in 2018, while showers more
intense than that of 2011 are not expected until 2023 and
2036. Possible returns of the shower produced by other perihelion
passages of the parent are also listed, though the absence of dust
production information for 3D/Biela means that such appearances are
more speculative. Nonetheless, careful observations of these showers
over the coming years could allow some measure of post-facto dust
measurement or even orbit improvement for this intriguing comet.

\acknowledgements{This work has received ongoing support from the
  Natural Sciences and Engineering Research Council of Canada and
  NASA's Meteoroid Environment Office.}

\bibliographystyle{natbib} 
\bibliography{Wiegert}

\newpage




\end{document}